\documentclass[pra,twocolumn,groupedaddress,superscriptaddress,showpacs,amsmath,amssymb,a4paper]{revtex4}
\usepackage{geometry}
\usepackage{graphicx}
\usepackage{verbatim}
\usepackage{float}
\usepackage{graphics}
\usepackage{mathrsfs}
\usepackage{amssymb}
\usepackage{amsmath}
\usepackage{amsthm}
\usepackage{bm}
\usepackage{hyperref}
\usepackage{braket}

\usepackage{amsbsy}

\theoremstyle{definition}

\newcommand{\be}{\begin{eqnarray}}
\newcommand{\ee}{\end{eqnarray}}

\addcontentsline{toc}{chapter}{References}

\begin{document}

\title{Dimension truncation for open quantum systems in terms of tensor networks}

\author{I. A. Luchnikov}

\affiliation{Moscow Institute of Physics and Technology,
Institutskii Per. 9, Dolgoprudny, Moscow Region 141700, Russia}

\affiliation{Skolkovo Institute of Science and Technology, Skolkovo Innovation Center, Building 3, Moscow
143026, Russia}

\author{S. V. Vintskevich }

\affiliation{Moscow Institute of Physics and Technology,
Institutskii Per. 9, Dolgoprudny, Moscow Region 141700, Russia}
\affiliation{A.M. Prokhorov General Physics Institute, Russian Academy of Science, Moscow, Russia}

\author{S. N. Filippov}

\affiliation{Moscow Institute of Physics and Technology,
Institutskii Per. 9, Dolgoprudny, Moscow Region 141700, Russia}

\affiliation{Institute of Physics and Technology, Russian Academy
of Sciences, Nakhimovskii Pr. 34, Moscow 117218, Russia}

\begin{abstract}
We present novel and simple estimation of a minimal dimension required for an effective reservoir in open quantum systems. Using a tensor network formalism we introduce a new object called  a reservoir network (RN). The reservoir network is the tensor network in the form of a Matrix Product State, which contains all effects of open dynamics. This object is especially useful for understanding memory effects. We discuss possible applications of the reservoir network and the estimation of dimension to develop new numerical and machine learning based methods for open quantum systems.

%We firstly give simple estimation of minimal dimension for an effective reservoir in open quantum systems using Tensor Network formalism. We introduce new object - reservoir network, which contains all effects of open dynamics. Reservoir network is the tensor in the Matrix Product State format. Especially, this object is useful to clearly understand the memory effects in open quantum systems. We discuss possible applications of reservoir network and dimension estimation to develop new numerical and machine learning based methods for open quantum systems.%

\end{abstract}


\maketitle

%----------------------------------------------------------------------
\section{Introduction}
One of the most challenging and interesting problems of modern theoretical physics is simulation of many-body quantum systems.
The dimension of Hilbert space grows exponentially with number of particles. This makes direct simulations impossible.
There are many analytical and numerical approaches, but all of them have limits of applicability. Very few models can be solved analytically, for example, using Bethe ansatz \cite{1}. Perturbation theory can be used only for systems with weak interactions. In case of systems with strong interaction, numerical approaches are very successful, but also have problems. For example, tensor networks and DMRG- based methods \cite{2,3,4,5,6,7,8} work well only for a $1d$ case, and cannot predict long-term time dynamics (because of Leib-Robinson boundary \cite{9}). Monte Carlo- based methods \cite{10} in fermionic case have sign problem. The many-body problem is still relevant and requires new approaches.
\\
We can divide this problem into two parts: stationary problem and dynamical problem. In this paper we focus on dynamic properties. In the case of real experiment we are very often interested only in the dynamics of small subsystem of the whole many-body system (open quantum systems dynamics). Problem of open quantum system dynamics \cite{11,12,13} without acceptance of Markov approximation is impossible to solve directly because of exponentially large dimension of a reservoir. In this work, we consider low-dimension reservoir approximation for an open quantum system. We provide a simple estimation of reservoir sufficient dimension in terms of \textbf{memory length, minimum time scale of reservoir, characteristic interaction constant} and \textbf{number of terms in the part of Hamiltonian describing interaction between system and reservoir}.
The most remarkable thing here that this estimation point to existence of low-dimension effective reservoirs. This allow us to perform simulations on classical computer using effective reservoirs. It is possible to develop new numerical and machine learning-based methods using low-dimension structure of reservoir in principle. The examples of successful application of the machine learning methods to many-body quantum physics can be found in \cite{14}.
Also we introduce a new object - \textbf{Reservoir Network}, which can be represented in Matrix Product State form \cite{2} or in Artificial Neural Network form \cite{14}. In the limit $\tau\rightarrow0$ this object is very similar to Continuous Matrix Product State \cite{21}. Finally, we estimate sufficient dimension of the reservoir applying tensor network formalism and providing estimation of an entangled entropy upper bound. At last, we can say that our work is another view of the Time-evolving block decimation algorithm \cite{15} at an angle of $90$ degrees.

\section{Tensor networks representation for quantum reservoir}
In this section we consider the most general model of open quantum system. The Hilbert space of the whole system plus reservoir ($S+R$) is ${\cal H}={\cal V}_S\otimes{\cal V}_R$. Usually, dimension of ${\cal V}_S$ is small, for instance it can be qbit space. On the other hand, a ${\cal V}_R$ is many-body quantum system with huge dimension. Naturally, it is assumed that dynamics of the whole system is completely dissipative, the memory length is finite and the Poincare recurrence time is infinite. All these assumptions lead to the Area law's analog in the condensed matter physics \cite{16,17} as shown further. The Hamiltonian of our system is $H=H_S\otimes I+I\otimes H_R + \gamma \sum_{i=1}^{n}A_i\otimes B_i$, where $\gamma$ is  a characteristic interaction rate. Let us divide Hamiltonian into two parts, $H_0=H_S\otimes I+I\otimes H_R$ and $H_{int}=\gamma \sum_{i=1}^{n}A_i\otimes B_i$ , with initial density matrix of whole system  $\rho(0)=\rho_S\otimes\rho_R$. Now let us consider the dynamical properties of whole system. Its evolution can be represented in terms of the Trotter decomposition:
\begin{eqnarray}
\rho(T)=\lim_{\tau\rightarrow 0}\ \underbrace{\Phi_0[\tau]\circ \Phi_{\rm int}[\tau]\circ \dots \circ\Phi_0[\tau]\circ \Phi_{\rm int}[\tau]}_{T/\tau}\rho(0).
\end{eqnarray}

\begin{figure}[H]
  \centering
  \includegraphics[scale=0.6]{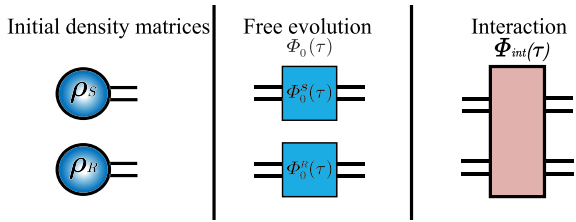}
  \caption{Key elements of tensor network representing system-reservoir dynamics:  initial density matrices, free-evolution  and interaction superoperators respectively.}   
  \label{fig1}
\end{figure}

Where $\Phi_0[\tau]=\exp[\tau {\cal L}_0]$, ${\cal L}_0\rho=-i[H_0,\rho]$ and $\Phi_{\rm int}[\tau]=\exp[\tau {\cal L}_{\rm int}]$, ${\cal L}_{\rm int}\rho=-i[H_{\rm int},\rho]$. Superoperators $\Phi_0$ and $\Phi_{\rm int}$ describe free evolution and interaction dynamics respectively in the the time scale $\tau$. We can illustrate all objects in  the Trotter decomposition in tensor networks terms (Figure \ref{fig1}).

\begin{figure}[H]
  \centering
  \includegraphics[scale=1.1]{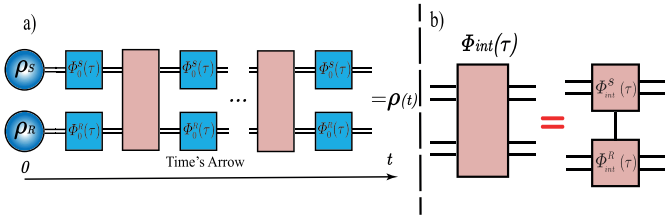}
  \caption{Tensor network. Edges represent convolutions.\\
  a) Tensor network representation of Trotter decomposition. \\
  b) The representation of $\Phi_{\rm int}(\tau)$ as a convolution of two tensors, details are shown in the Appendix 1.}   
  \label{fig2}
\end{figure}

Trotter decomposition in this terms takes the following form (Figure \ref{fig2}).

\begin{figure}[H]
    \centering
    \includegraphics[scale=0.6]{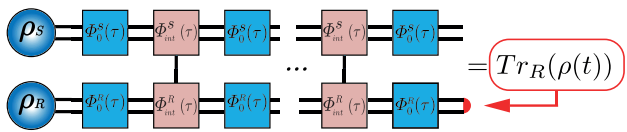}
    \caption{Tensor networks representation of partial density matrix of system. A red loop on the right side represents trace taken over reservoir.}
    \label{fig3}
\end{figure}

\begin{figure}[H]
    \centering
    \includegraphics[scale=0.6]{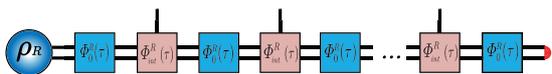}
    \caption{Reservoir network. It is exactly the same as a Matrix Product State network. (an analog of $1{\rm d}$ quantum chains).}
    \label{fig4}
\end{figure}

\begin{figure*}
    \centering
    \includegraphics[scale=5]{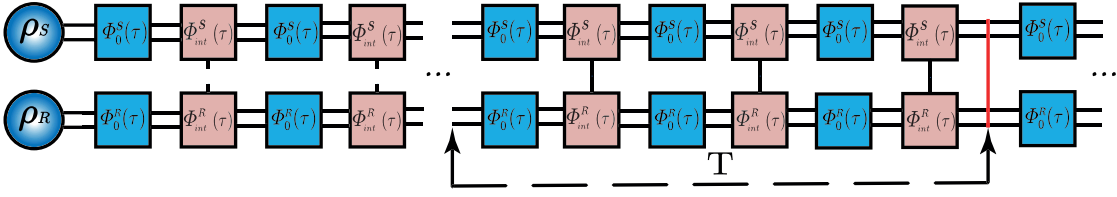}
    \caption{Partial density matrix for reservoir network. Red line shows boundary which cut far part (part that is behind the memory length).}
    \label{fig5}
\end{figure*}

Finally, the partial density matrix of the system takes the following form (Figure \ref{fig3}).
At this stage we introduce new interesting and important object - \textbf{Reservoir Network (RN)}. The RN  is schematically represented in (Figure \ref{fig4}). It contains all information about quantum reservoir and about all open dynamics effects (memory effects, dissipation, etc.), and is obtained by cutting tensor network horizontally where the lower part is another tensor network.
This object is exactly the same as the Matrix Product State and we can treat this object like a quantum state \cite{2}. Furthermore, the analog of a density matrix and partial density matrix for this object can be constructed in terms of tensor networks. %The entangled entropy is another important property which can be defined for this object.%
Also, let us define the connection between a memory length and a mutual information for the RN. It is known that
mutual information between two subsystems with density matrices $\rho_{A}$ and $\rho_{B}$ of a quantum system in non-critical $1d$ case satisfies an inequality $I(\rho_A,\rho_B)\leq C\exp\Big[-\frac{l}{L}\Big]$. Here $l$ is number of vertexes between $A$ and $B$ subsystems and $L$ is  a correlation length. $I(\rho_A,\rho_B)=S(\rho_A)+S(\rho_B)-S(\rho_{AB})$.
In our case correlation length $L$ is replaced by characteristic memory length. In this case, we can expect behavior similar to the case of $1{\rm d}$ low energy quantum chains with a local Hamiltonian. We can expect that there is an MPS approximation with low dimension auxiliary space (in our case auxiliary space is reservoir space). For a better understanding connection between the RN and the MPS is presented in the Table 1.
\begin{table}[h!]
     \centering
    \begin{tabular}{p{4.2cm}|p{4.2cm}}
    Reservoir Network &  Matrix Product State\\
    \hline \\
   Time & Spatial coordinate
\\\hline\\
       Memory length  & Correlation length\\\hline\\
      Dimension fo reservoir & Dimension of auxiliary space
      \\ \hline \\
     $n$ from $H_{\rm int}=\gamma \sum_{i=1}^{n}$\\$A_i\otimes B_i$ (see Appendix 1) & Dimension of chain subsystem
     \\ \hline \\
     Area law (because memory length is finite) & Area law
     \\ \hline \\
     Mutual Information  &  Entangled entropy
    \end{tabular}
    \caption{Comparison between the Matrix Product State and the Reservoir Network.}
    \label{tab:my_label}
\end{table}
\\ Now let us do a  non-formal estimation of sufficient dimension for auxiliary space. It is known that the MPS with small entangled entropy can be truncated. Let us give a remind how it can be done, for example, in the condensed matter physics and quantum information \cite{18,19}. Suppose we have two component quantum system. Wave function of this system takes the following form $\ket{\psi}=\sum_{i,j}\Psi_{i,j}\ket{i}\otimes\ket{j}$. We can do Singular Value Decomposition (SVD) for $\Psi_{i,j}$ matrix.
$\ket{\psi}=\sum_{i,j,\alpha}U_{i \alpha}\lambda_\alpha V^\dagger_{\alpha j}\ket{i}\otimes\ket{j}=\sum_\alpha \lambda_\alpha \ket{\alpha}\otimes\ket{\alpha}$, which is also known as the Schmidt decomposition. Number of non-zero singular values is called the Schmidt rank. We can truncate Schmidt rank neglecting the minimum singular values (optimal low-rank approximation). Let us define the new small Schmidt rank by considering partial density matrices for subsystems.  $\rho_1=\sum_{j}\lambda_\alpha^2\ket{\alpha}\bra{\alpha}$ and $\rho_2=\sum_{j}\lambda_\alpha^2\ket{\alpha}\bra{\alpha}$ density matrices for the first and second subsystems respectively. Entangled entropy is $S(\rho_1)=-\sum_j\lambda_j\log\lambda_j$. For whole system with the Schmidt rank $d$, the maximal entangled entropy is defined by $S_{\rm max}=\log d$. Here we emphasize that a naturally sufficient dimension takes following form $\log(d_{\rm suff})=S$. The trick is that the Schmidt rank describes only the entangled entropy less or equal then $S$.  
As a result, for an auxiliary space, sufficient dimension is given by $d_{\rm suff}=\exp S$. We use this trick in Appendix $2$ for more precise sufficient dimension estimation. 
\\
On the next step we consider \textbf{Reservoir Partial Density Matrix (RPDM)} for our RN (Fig.\ref{fig5}). Notice that ${\rm Number\ of\ connections} \propto T$ in the considered PRDM and $d_{\rm suff}\propto\exp[{\rm Number\ of\ connections}]$. Finally, $d_{\rm suff}\propto\exp[\gamma T]$, where $\gamma$ appears from simple physical dimensional analysis. More precise estimation is:
\begin{eqnarray}
d_{\rm suff}\approx\exp\left(2n\gamma T\left[1-\log\left(\gamma t\right)\right]\right).
\end{eqnarray}
The derivation is presented in the Appendix 2.

\section{Discussion}
Based on analogy between Matrix Product State and time decomposition of reservoir we developed new language for open quantum systems (OQS) which can be effectively applied to its  numerical analysis. For instance, we showed existence of valid low dimension approximation of quantum reservoir for any open quantum systems with finite memory length. Let us consider a simple example to understand characteristic dimension of effective low-dimension reservoir. Suppose, that $\frac{1}{\gamma}\approx T$, $n=2$ and $t\approx\frac{1}{5\gamma}$. From eq. (2) one can find that $d_{\rm suff}\approx1000$. As one can see, it is a reservoir containing just $10$ qbits and can be easily simulated on a computer. Using this fact, we can develop innovative numerical and machine learning based  methods of research for the OQS. For example, choosing ensemble of preliminary measurements, one can reconstruct a structure of quantum reservoir. Such reconstruction procedure is possible because the low dimension approximation is described by a small number of parameters. This technique is very similar to the Hidden Markov Models in the voice recognition problem \cite{20}. On the other hand, we can try to use variation principle and reconstruct a reservoir using optimization algorithms. The considered approximation becomes similar to optimization based reconstruction procedure of the MPS in the condensed matter physics. We suggest a new family of objects (RN and etc.) which can provide a convenient and useful formalism for open quantum systems. Furthermore, existence of such approximation is an interesting fundamental result. 
\section{Appendix 1}
Here we present some detailed explanations of RN definition. The Hamiltonian of our system is given by
\begin{eqnarray}
H=H_{S}\otimes I+I\otimes H_{R}+\gamma\sum^n_{i=1} A_i \otimes B_i.
\label{0}
\end{eqnarray}
For better understanding let us introduce the new notation. First, we represent the density matrix in the column form:
\begin{eqnarray}
&&\varrho=\sum_{i,j}\rho_{i,j}\ket{i}\bra{j}\ \rightarrow \ \ket{\varrho}=\sum_{i,j}\rho_{i,j}\ket{i}\otimes \ket{j}, \nonumber \\
&&A\varrho B\ \rightarrow \ A\otimes B^T \ket{\varrho}.
\end{eqnarray}
Second, the time dynamics of density matrix at time $\tau$ takes the following form:
\begin{eqnarray}
\ket{\varrho(\tau)}=\exp[-i\tau H]\otimes \exp[i\tau H^*]\ket{\varrho}.
\end{eqnarray}
 Using the Baker Hausdorff equality and assuming that $\tau$ is small, one can obtain the following expression:
\begin{eqnarray}
&&\exp[-i\tau H]\otimes \exp[i\tau H^*]=\nonumber\\&&=\exp[-i\tau H_{0}]\otimes\nonumber\\&&\otimes \exp[i\tau H_{0}^*]\exp[-i\tau H_{\rm int}]\otimes \exp[i\tau H_{\rm int}^*]+O(\tau^2).
\end{eqnarray}
Where $H_0=H_{S}\otimes I+I\otimes H_{R}$ and $H_{\rm int}=\gamma\sum^n_{i=1} A_i \otimes B_i$. Here we divided dynamics of whole system into two parts
\begin{eqnarray}
\Phi_0=\exp[-i\tau H_{0}]\otimes \exp[i\tau H_{0}^*].
\end{eqnarray}
\begin{eqnarray}
&&\Phi_{\rm int}(\tau)=\exp[-i\tau H_{\rm int}]\otimes \exp[i\tau H_{\rm int}^*]=\nonumber\\&&=I\otimes I \otimes I \otimes I -i\tau\gamma\sum_{i=1}^n A_i\otimes B_i \otimes I \otimes I+\nonumber\\&&+i\tau\gamma\sum_{i=1}^n  I \otimes I \otimes A_i^*\otimes B_i^*+O(\tau^2).
\end{eqnarray}
For convenience, we rearrange the Hilbert spaces:
\begin{eqnarray}
&&\Phi_{\rm int}(\tau)=I\otimes I \otimes I \otimes I -i\tau\gamma\sum_{i=1}^n A_i\otimes I \otimes B_i \otimes I+\nonumber\\&&+i\tau\gamma\sum_{i=1}^n  I \otimes A_i^* \otimes I \otimes B_i^*+O(\tau^2).
\end{eqnarray}
Using this notation we can rewrite it in the short form:
\begin{eqnarray}
\Phi_{\rm int}(\tau)=\sum_{i=0}^{2n} {\cal A}_i \otimes {\cal B}_i.
\end{eqnarray}
Where,
\begin{eqnarray}
&&{\cal A}_i=\begin{cases} I\otimes I,\ i=0 \\\sqrt{\gamma\tau}A_i \otimes I,\ 0<i<n+1 \\ \sqrt{\gamma\tau}I\otimes A_{i-n}^*,\ n<i,   \end{cases}\nonumber\\
&&{\cal B}_i=\begin{cases} I\otimes I,\ i=0 \\-i\sqrt{\gamma\tau} B_i \otimes I,\ 0<i<n+1 \\i\sqrt{\gamma\tau} I\otimes B_{i-n}^*,\ n<i.   \end{cases}
\label{1}
\end{eqnarray}
Such separation of $\gamma\tau$ is non-trivial. In order to explain this separation we consider a two point correlation function for the RN. This function is well-defined only for presented separation for $\gamma\tau \rightarrow \sqrt{\gamma\tau}$ and $\sqrt{\gamma\tau}$.

\begin{figure}
    \centering
    \includegraphics[scale=0.45]{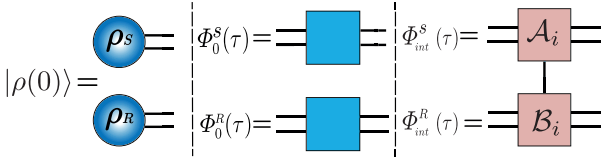}
    \caption{Reservoir network. Notice that superscripts $S$ and $R$ correspond to upper and lower parts of the network respectively. For simplicity of diagrams, these superscripts are omitted further.}
    \label{fig6}
\end{figure}

\begin{figure}
    \centering
    \includegraphics[scale=0.6]{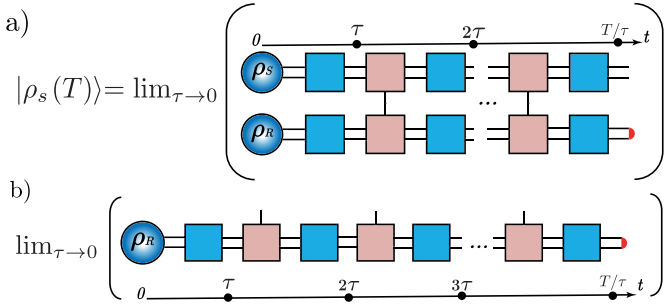}
    \caption{Density matrix of system $+$ reservoir in terms of the Tensor Network and the Reservoir Network. \\
    a) Tensor network representation for $\rho_{\rm S}(T)$.\\
    b) Tensor network representation for quantum reservoir. This object is actually multidimensional tensor in the MPS form. Any time marks  $\tau, 2\tau, 3\tau$ etc. are equivalent to the network nodes of the MPS, whereas full number of nodes equals $T/\tau$.}
    \label{fig7}
\end{figure}

\begin{figure}
    \centering
    \includegraphics[scale=0.65]{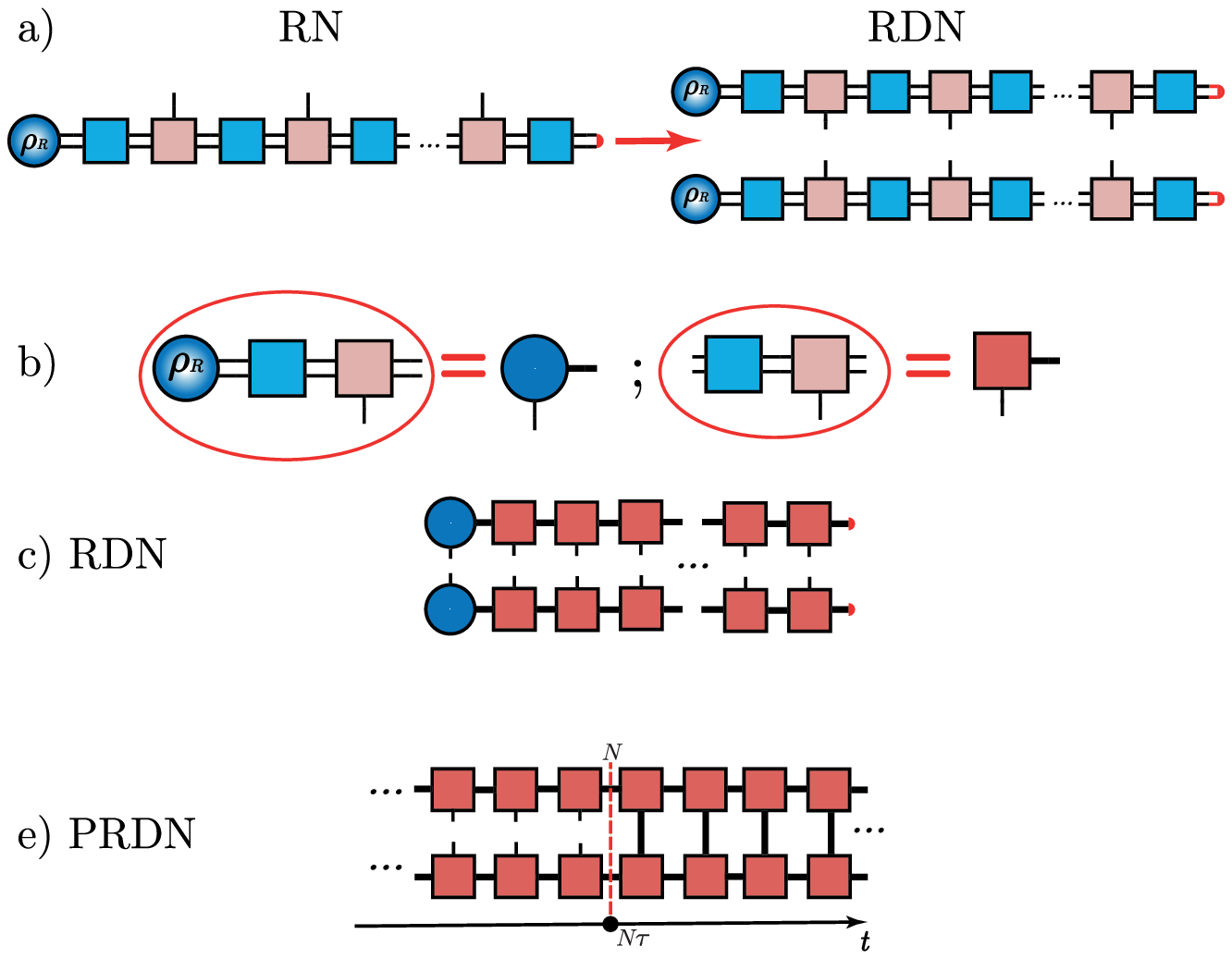}
    \caption{Density matrix of system in terms of the Tensor Network and the Reservoir Network (RN). Diagrammatic construction of Reservoir Density Network (RDN) and Partial Reservoir Density Network. \\
    a) Reservoir network (RN) in terms of the MPS and Reservoir density network (RDN)\\ 
    b) The replacement and simplification of nodes of in the RDN \\
    c) The RDN after simplification \\
    d) The Partial Reservoir Density Network (PRDN).  Connected indeces between upper and lower parts of network mean partial trace in selected subsystem, which starts from some node with number N.}
    \label{fig8}
\end{figure}

Our goal is to represent the whole system evolution in terms of tensor networks (Fig\ref{fig6}).
We can give diagrammatic formulation for the density matrix of the system in terms of tensor network diagrams as well as for the RN. (Fig\ref{fig7}).
Owing to Reservoir Network (RN) can be interpreted as a quantum state (formal analogy: RN $\rightarrow$ $\ket{\psi}$) one can discern a generalization of this analogy further. Based on the RN new tensor network can be easily constructed, which we may treat as a reservoir density tensor network (RDN) in accordance with complete analogy of a density matrix. This idea is illustrated in the (Fig.\ref{fig8}).

\section{Appendix 2}
\begin{figure}
    \centering
    \includegraphics[scale=0.6]{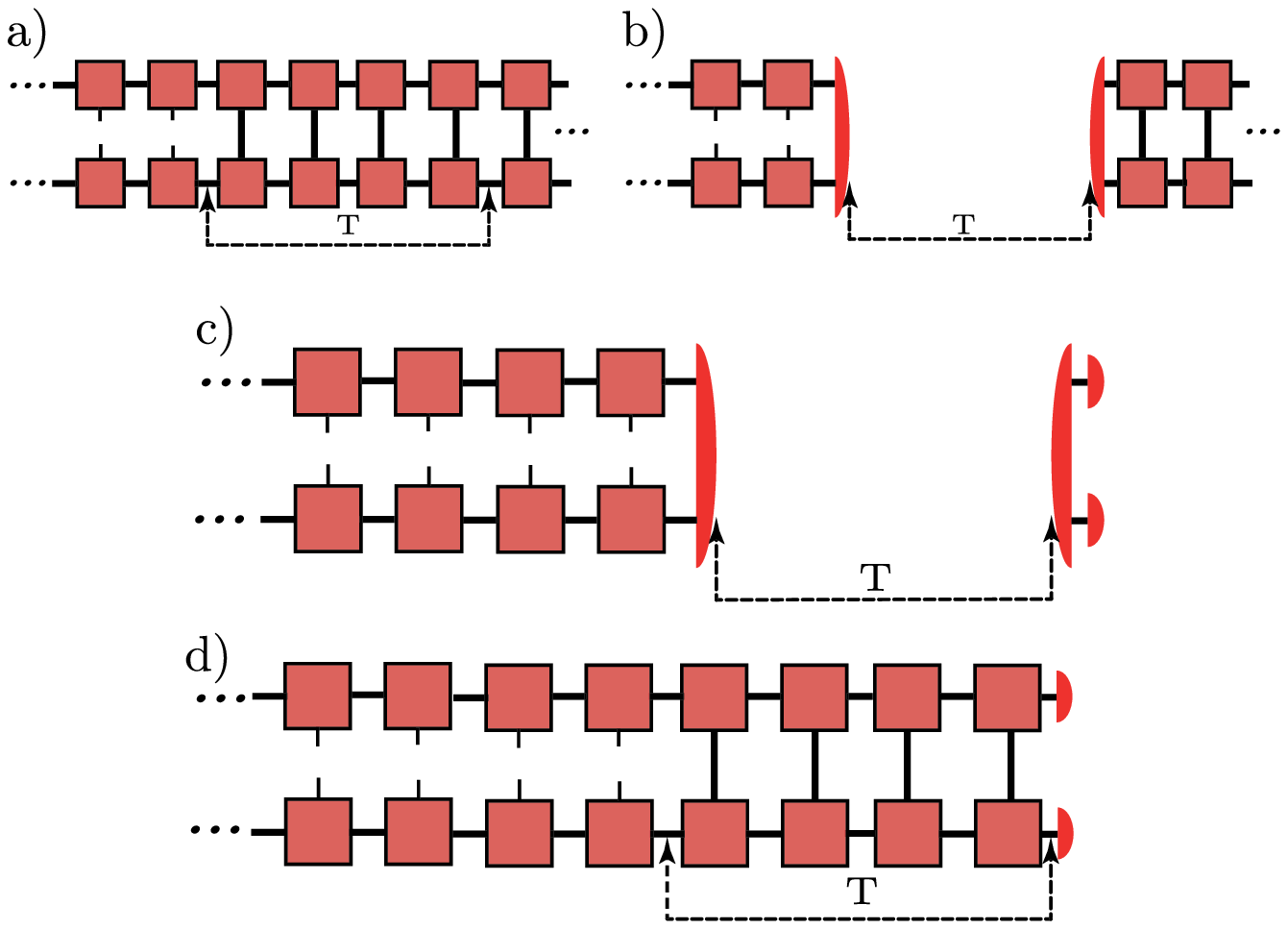}
    \caption{Chain of transformations for RDN.\\ 
    a) A selection subsystem of reservoir with some memory time $T$.\\
    b) Using assumption $I(\rho_A,\rho_B) \approx 0 $ for finite $T$, inner part can be replaced by ancillary tensor network without this inner part. The ancillary network doesn't have any connection between remaining parts from both sides of the replaced one. \\
    c) The next step we stress that the right part of the network obtained from previous step b) essentially is a constant, which can be represented as an arbitrary convolution. \\ 
    d) Previous steps show that inner part can be replaced to the tensor without internal convolutions and vice versa. Such opportunity occurs because of absence of any correlations between parts on sides from inner one.}  
    \label{fig9}
\end{figure}

\begin{figure}
    \centering
    \includegraphics[scale=0.8]{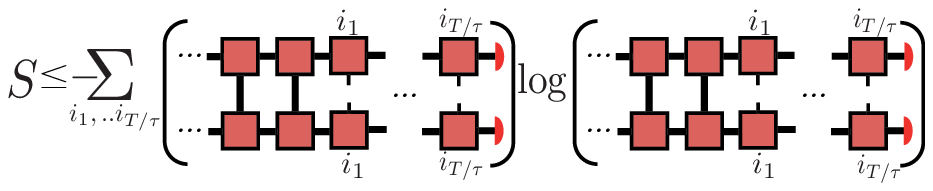}
    \caption{Entropy estimation within diagrammatic representation.}
    \label{fig10}
\end{figure}

\begin{figure}
    \centering
    \includegraphics[scale=0.65]{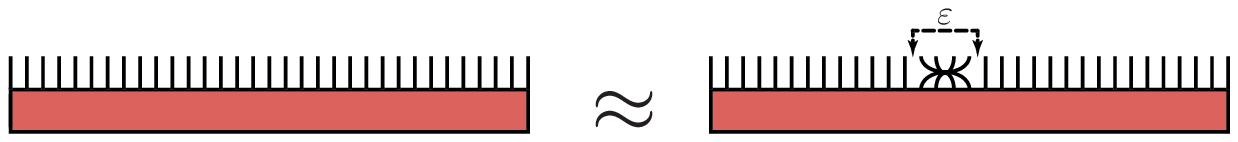}
    \caption{Replacement indices in small time-neighbourhood of RN doesn't change it much}
    \label{fig11}
\end{figure}

The Appendix shows the estimation of entangled entropy. Let us start with simple example: consider an arbitrary density matrix decomposition on the \textbf {not orthogonal} and \textbf {not normalize} set $\{\ket{j}\}$. In accordance with previous discussion more precise estimation of entangled entropy needs to be done. In our case for $\rho=\sum_j \ket{j}\bra{j}$ the entropy of this density matrix is $S\leq-\sum_j \braket{j|j}\log \braket{j|j}$.
Let us prove this inequality. We can rewrite object $\ket{j}\bra{j}$ in the following form:
\begin{eqnarray}
\ket{j}\bra{j}=q_jP_j,
\end{eqnarray}
where $P_j^2=P_j$ and $q_j=\braket{j|j}$. For the Renyi entropy with $\alpha\geq1$ we have:
\begin{eqnarray}
&&S_\alpha=\frac{1}{1-\alpha}\log\left[{\rm tr}\left(\sum_j q_jP_j\right)^\alpha\right]=\nonumber\\&&=\frac{1}{1-\alpha}\log\left[{\rm tr}\left(\sum_j q_jP_j\right)^\frac{\alpha}{2}\left(\sum_j q_jP_j\right)^\frac{\alpha}{2}\right]=\nonumber\\&&=
\frac{1}{1-\alpha}\log\left[\Big\|\sum_j q_jP_j\Big\|_\alpha^\alpha\right]\leq\nonumber\\&&\leq\frac{1}{1-\alpha}\log\left[\Big(\sum_j q_j\|P_j\|_\alpha\Big)^\alpha\right]\leq\nonumber\\&&\leq\frac{1}{1-\alpha}\log\left[\sum_j q_j^\alpha\right]
\end{eqnarray}
For $\alpha=1$ (Von Neumann entropy) the inequality is proved.
Similarly we estimate upper bound of the entangled entropy for our PRDM, where the PRDM is assumed to be normalized. Following chain of transformations leads to cut of PRDM's tail. Thus procedure is presented on (Fig\ref{fig9}). The whole chain of transformation based on inequality $I(\rho_A,\rho_B)\leq C\exp\Big[-\frac{l}{T}\Big]$.Here $T$ is memory length, l is number of blocks between $A$ and $B$ subsystems, $C$ is just some constant.
Finally, using $S\leq-\sum_j \braket{j|j}\log \braket{j|j}$ we can represent entropy estimation in the diagrammatic form Fig.(\ref{fig10}).
Estimation of logarithmic term :
\begin{eqnarray}
&&{\cal A}_i=\begin{cases} I\otimes I,\ i=0 \\\sqrt{\gamma\tau}A_i \otimes I,\ 0<i<n+1 \\ \sqrt{\gamma\tau}I\otimes A_{i-n}^*,\ n<i,   \end{cases}\nonumber\\
&&{\cal B}_i=\begin{cases} I\otimes I,\ i=0 \\-i\sqrt{\gamma\tau} B_i \otimes I,\ 0<i<n+1 \\i\sqrt{\gamma\tau} I\otimes B_{i-n}^*,\ n<i.   \end{cases}
\label{1}
\end{eqnarray}
We can state (see Appendix 1)
\begin{eqnarray}
{\rm Norm}_{i_1,i_2,\dots,i_{T/\tau}}\approx N (\gamma\tau)^{I(i_1)+I(i_2)+\dots+I(i_{T/\tau})},
\end{eqnarray}
where ${\rm Norm_{i_1,i_2,\dots,i_{T/\tau}}}$ is the term under logarithm, $N$ is some constant and \\$I(i)=\begin{cases}1,\ \ i\neq0\\0,\ \ i=0\end{cases}$. Now we should derive constant $N$. We can use normalization condition:
\begin{eqnarray}
1=\sum_{i_1,i_2,\dots,i_{T/\tau}}{\rm Norm}_{i_1,i_2,\dots,i_{T/\tau}}=N(1+2n\gamma\tau)^{T/\tau}.
\end{eqnarray}
Now we can find entropy estimation
\begin{eqnarray}
&&S\approx -\frac{2n\gamma T}{1+2n\gamma\tau}\log\frac{\gamma\tau}{1+2n\gamma\tau}-\nonumber\\&&-\frac{1}{1+2n\gamma\tau}\log\frac{1}{1+2n\gamma\tau}
\end{eqnarray}
For the small $\gamma\tau$ we have
\begin{eqnarray}
S\approx2n\gamma T\left[1-\log\left(\gamma\tau\right)\right]
\end{eqnarray}
Unfortunately, this expression diverges in the limit $\tau\rightarrow0$. However, divergence has simple explanation  and can be removed. This problem is similar to the problem of the divergence of the classical ideal gas entropy. The solution based on fact that the minimal phase volume for the particle state is equal $(2\pi\hbar)^3$. Hence the entropy of the classical ideal gas becomes finite. But what is the analog of this phase volume in this particular case? Let us consider some $\epsilon$ neighborhood in the time axis. If $\epsilon$ is small enough, we can replace any physical indices in this $\epsilon$ neighborhood Fig(\ref{fig11}).

 Thus, $\epsilon$ neighborhood is a direct analog of minimal phase volume for the gas. What is the physical meaning of this? We can say that this is \textbf{minimal time scale for the reservoir} - $\tau_{\rm min}$. Using this fact we can replace $\tau$ to $\tau_{\rm min}$ in our expressions and divergence consequently disappears . Finally, taken previous assumptions into account for the expression of the entropy we have:
\begin{eqnarray}
S\approx2n\gamma T\left[1-\log\left(\gamma t\right)\right].
\end{eqnarray}
As a main result the sufficient dimension of reservoir is given by:
\begin{eqnarray}
d_{\rm suff}\approx\exp\left(2n\gamma T\left[1-\log\left(\gamma t\right)\right]\right).
\end{eqnarray}

\end{document}